\documentclass[article, 10pt, twocolumn, twoside]{IEEEtran}

\usepackage{theorem}
\usepackage{amsmath}
\usepackage{amssymb}
\usepackage{graphicx}
\usepackage[list=off]{caption}
\usepackage{enumitem}
\usepackage[noadjust]{cite}
\usepackage[english]{babel}
\usepackage{epsfig,amssymb,amsfonts}
\usepackage{theorem}
\usepackage{amsmath}
\usepackage{amssymb}
\usepackage{bbm}
\usepackage{cite}
\usepackage{subfigure}
\usepackage{tikz}
\usepackage{tkz-berge}
\usepackage{pgfplots}
\usetikzlibrary{plotmarks}
\usetikzlibrary {positioning}

\def\Ab{\mathbf{A}}
\def\Bb{\mathbf{B}}
\def\Cb{\mathbf{C}}
\def\Db{\mathbf{D}}

\def\Iden{\mathbf{I}}

\def\Kb{\mathbf{K}}
\def\Mb{\mathbf{M}}

\def\Pb{\mathbf{P}}
\def\Vb{\mathbf{V}}

\def\yb{\mathbf{y}}
\def\xb{\mathbf{x}}
\def\zb{\mathbf{z}}
\def\ub{\mathbf{u}}

\def\nb{\mathbf{n}}

\def\Sig{\boldsymbol{\Sigma}}
\def\Omegab{\boldsymbol{\Omega}}

\def\mub{\boldsymbol{\mu}}
\def\0b{\boldsymbol{0}}

\def\cg{\left[}
\def\cd{\right]}
\def\vg{\left\{}
\def\vd{\right\}}
\def\pg{\left(}
\def\pd{\right)}

\def\vec{\textup{vec}}

\newtheorem{Thm}{Theorem}[section]
\newtheorem{remark}{Remark}[section]
\newtheorem{corollary}{Corollary}[section]

\newenvironment{myproof}[1][\proofname]{\IEEEproof[#1]}{\endIEEEproof}

\begin{document}
\title{New insights into the statistical properties of $M$-estimators}


\author{Gordana Dra\v{s}kovi\'{c},~\IEEEmembership{Student Member,~IEEE}, Fr\'{e}d\'{e}ric Pascal,~\IEEEmembership{Senior Member,~IEEE}\thanks{Gordana Dra\v{s}kovi\'{c} and Fr\'{e}d\'{e}ric Pascal are with L2S - CentraleSup{\'e}lec - CNRS - Universit\'{e} Paris-Sud - 3 rue Joliot-Curie, F-91192 Gif-sur-Yvette Cedex, France (e-mails: gordana.draskovic@l2s.centralesupelec.fr, frederic.pascal@l2s.centralesupelec.fr). ``This paper has supplementary downloadable material available at http://ieeexplore.ieee.org, provided by the author. The material includes the results for the real case. This material is 149KB in size.''}}

\maketitle

\begin{abstract}
This paper proposes an original approach to better understanding the behavior of robust scatter matrix $M$-estimators. Scatter matrices are of particular interest for many signal processing applications since the resulting performance strongly relies on the quality of the matrix estimation. In this context, $M$-estimators appear as very interesting candidates, mainly due to their flexibility to the statistical model and their robustness to outliers and/or missing data. However, the behavior of such estimators still remains unclear and not well understood since they are described by fixed-point equations that make their statistical analysis very difficult. To fill this gap, the main contribution of this work is to prove that these estimators distribution is more accurately described by a Wishart distribution than by the classical asymptotical Gaussian approximation. To that end, we propose a new ``Gaussian-core'' representation for Complex Elliptically Symmetric (CES) distributions and we analyze the proximity between $M$-estimators and a Gaussian-based Sample Covariance Matrix (SCM), unobservable in practice and playing only a theoretical role. To confirm our claims we also provide results for a widely used function of $M$-estimators, the Mahalanobis distance. Finally, Monte Carlo simulations for various scenarios are presented to validate theoretical results.  
\end{abstract}


\section{Introduction}
In signal processing applications, the knowledge of scatter matrix is of crucial importance. It arises in diverse applications such as filtering, detection, estimation or classification. In recent years, there has been growing interest in covariance matrix estimation in a vast amount of literature on this topic (see e.g., \cite{ollila2012complex, pascal2008covariance, chen2011robust, wiesel2012unified, pascal2012mggd, mahot2013asymptotic, Palomar-14, ollila2014regularized} and references therein). Generally, in most of signal processing methods the data can be locally modelled by a multivariate zero-mean circular Gaussian stochastic process, which is completely determined by its covariance matrix. Complex multivariate Gaussian, also called complex normal (CN), distribution plays a vital role in the theory of statistical analysis \cite{gupta2000matrix}. Very often the multivariate observations are approximately normally distributed. This approximation is (asymptotically) valid even when the original data is not multivariate normal, due to the central limit theorem. In that case, the classical covariance matrix estimator is the sample covariance matrix (SCM) whose behavior is perfectly known. Indeed, it follows the Wishart distribution \cite{statistics1999} which is the multivariate extension of the gamma distribution. Thanks to its explicit form, the SCM is easy to manipulate and therefore widely used in the signal processing community.

Nevertheless, the complex normality sometimes presents a poor approximation of underlying physics. Noise and interference can be spiky and impulsive i.e., have heavier tails than the Gaussian distribution. An alternative has been proposed by introducing elliptical distributions \cite{kelker1970distribution}, namely the Complex Elliptically Symmetric (CES) distributions. These distributions present an important property which states that their higher order moment matrices are scalars multiple of their correspondent normal distribution. This presents a starting point for the analysis that is done in this paper. These distributions have been frequently employed for non-Gaussian modeling (see e.g., for radar applications \cite{Gini99b, Gini00c, Gini02, Conte87, Conte02b}).

Although Huber introduced robust $M$-estimators in \cite{huber1964robust} for the scalar case, Maronna provided the detailed analysis of the corresponding scatter matrix estimators in the multivariate real case in his seminal work \cite{Maronna76}. $M$-estimators correspond to a generalization of the well-known Maximum Likelihood estimators (MLE), that have been widely studied in the statistics literature \cite{kent1988mleCauchy,balleri2007mle}. In contrast to $ML$-estimators where the estimating equation depends on the probability density function (PDF) of a particular CES distribution, the weight function in the $M$-estimating equation can be completely independent of the data distribution. Consequently, $M$-estimators presents a wide class of scatter matrix estimators, including the $ML$-estimators, robust to the data model. In \cite{Maronna76}, it is shown that, under some mild assumptions, the estimator is defined as the unique solution of a fixed-point equation and that the robust estimator converges almost surely (a.s.) to a deterministic matrix, equal to the scatter matrix up to a scale quantity (depending on the true statistical model). Their asymptotical properties have been studied by Tyler in the real case \cite{tyler1982radial}. This has been recently extended to the complex case, more useful for signal processing applications, in \cite{ollila2012complex, mahot2013asymptotic}.

In most of the papers, three main $M$-estimators are studied and used in practice: the Student's $M$-estimator that is MLE for $t$-distribution, the Huber's $M$-estimator and the Tyler's $M$-estimator \cite{tyler1987distribution}, also known as Fixed Point (FP) estimator \cite{pascal2008covariance}. Student $t$-distribution is widely employed for non-Gaussian data modeling since it offers flexibility thanks to an additive parameter, namely the Degree of Freedom (DoF). As a consequence, Student's $M$-estimator is often used for scatter matrix estimation. Huber's $M$-estimator, especially its complex multivariate extension, has received a lot of attention since proven to be very robust to outliers. Tyler's $M$-estimator is not exactly an $M$-estimator\footnote{especially because it does not respect all Maronna conditions \cite{Maronna76}} but it is very useful because of rare property that any CES distribution with the same scatter matrix leads to the same result (hence ``distribution-free''). Asymptotical properties of this estimator have been analyzed in \cite{pascal2008performance, ollila2012complex}. Recently, it has been shown that the behavior of Tyler's estimator can be better approximated by a Wishart distribution \cite{draskovic2016TyE}. In this work, one aims at providing more general results that can be applied to all $M$-estimators and one wants to analyze the gain of this approach on the robust Mahalanobis distance \cite{mahalanobis1936generalized,DEMAESSCHALCK20001}, very useful in various problems such as detection, clustering etc.

The contributions of this work are multiple. First, the originality of the results comes from a new CES representation introducing ``Gaussian cores''. This representation is a modified stochastic representation given in \cite{ollila2012complex} and is crucial to understand the proposed method.
Second, in this paper, $M$-estimators are, for the first time, analyzed thanks to a comparison with a very simple estimator, the SCM. Indeed, the direct statistical analysis of these estimators is difficult because they are defined as the solution of an implicit equation and have been analyzed only in classical asymptotic regimes. Here, we propose a different approach to overcome this difficulty. More precisely, a sort of distance between $M$-estimators and the SCM is computed in order to propagate SCM non-asymptotic properties  towards $M$-estimators. Third, the paper gives new insights into the correlation between $M$-estimators and the corresponding SCM in the Gaussian context which is the central part of our approach. Finally, we present a practical interest of the results, specifically the application to the Mahalanobis distance. Note that all the results are provided in the complex case. For completeness purposes, a supplemental material containing analogous results in the real case is provided together with this article.

The rest of this paper is organized as follows. Section \ref{sec:problem} introduces the considered CES-models based on Gaussian cores as well as the $M$-estimators and Mahalanobis distance. Section \ref{sec:contrib} contains the main contribution of the paper with discussions and further explanations. Moreover, closed-form expressions are derived for some particular cases of $M$-estimators and the application to the Mahalanobis distance is presented. In Section \ref{sec:simus}, Monte Carlo simulations are presented in order to validate the theoretical results. Finally, some conclusions and perspectives are drawn in Section \ref{sec:conclu}.\\

\textit{Notations} - Vectors (resp. matrices) are denoted by bold-faced lowercase letters (resp. uppercase letters). $^T$, $^*$ $^H$ respectively represent the transpose, conjugate and the Hermitian operator. $Re(.)$ and $Im(.)$ denote respectively the real and the imaginary part of a complex quantity, i.i.d. stands for ``independent and identically distributed'' while $\sim$ means ``is distributed as''. $=_d$ stands for ``shares the same distribution as'', $\overset{d}{\to}$ denotes convergence in distribution and $\otimes$ denotes the Kronecker product. $\mathbbm{1}$ is indicator function and $\vec(\cdot)$ is the operator which transforms a matrix $ m \times n$ into a vector of length $mn$, concatenating its $n$ columns into a single column. Moreover, $\Iden_m$ is the $m \times m$ identity matrix, $\0b$ the matrix of zeros with appropriate dimension and $\Kb$ is the commutation matrix (square matrix with appropriate dimensions) which transforms $\vec(\Ab)$ into $\vec(\Ab^T)$, i.e. $\Kb\,\vec(\Ab) = \vec(\Ab^T)$. 

\section{Problem formulation}
\label{sec:problem}

\subsection{Complex distributions}
Let $\zb=Re(\zb)+jIm(\zb)$ be an $m$-dimensional complex random vector which consists of a pair of real random vectors $Re(\zb)$ and $Im(\zb)$. The distribution of $\zb$ on $\mathbb{C}^m$ determines the joint real $2m$-variate distribution of $Re(\zb)$ and $Im(\zb)$ on $\mathbb{R}^{2m}$ and conversely. To completely define the second-order moments of $Re(\zb)$ and $Im(\zb)$, $\zb$ is given by its covariance matrix $\Cb=E[(\zb-\mub)(\zb-\mub)^H]$ and pseudo-covariance matrix $\Pb=E[(\zb-\mub)(\zb-\mub)^T]$. If the complex vector is \textit{circular} (see \cite{ollila2012complex} for details), the pseudo-covariance vanishes, i.e. $\Pb=\0b$. 

\subsubsection{Generalized Complex Normal distribution}
An $m$-dimensional random vector has the generalized normal distribution $\zb\sim\mathbb{C}N(\mub,\Cb,\Pb)$ if its probability density function (PDF) can be written as
\begin{equation}
\label{pdfgn}
h_{\zb}(\zb)=\frac{\exp{\vg-\frac{1}{2}\cg(\zb-\mub)^H \quad (\zb-\mub)^T\cd \Vb^{-1} \cg  \begin{matrix} \zb-\mub \\ \zb^*-\mub^* \end{matrix}  \cd\vd}}{\pi^m \sqrt{|\Vb|}}
\end{equation}
where $\mub$ is the statistical mean and $\Vb=\cg\begin{matrix} \Cb & \Pb \\ \Pb^* & \Cb^* \end{matrix}\cd$.
If $\zb$ is circular CN-distributed the pseudo-covariance will be omitted in the notation, i.e. $\zb\sim\mathbb{C}N(\mub,\Cb)$.

\subsection{Complex Elliptically Symmetric distributions}
An important class of circular distributions are the CES distributions.
An $m$-dimensional random vector has a CES distribution if its probability density function (PDF) can be written as
\begin{equation}
\label{pdf}
h_{\zb}(\zb)=C |\Mb |^{-1}\,g_{\zb}\pg (\zb-\mub)^H \Mb^{-1}(\zb-\mub)\pd
\end{equation} 
where $C$ is a constant, $g_{\zb}:[0,\infty)\rightarrow [0,\infty)$ is any function (called the density generator) such that \eqref{pdf}  defines a PDF  and $\Mb$ is the scatter matrix. The matrix $\Mb$ reflects the structure of the covariance matrix $\Cb$ of $\zb$, i.e., the covariance matrix is equal to $\Mb$ up to a scale factor $\Cb=\xi\Mb$\footnote{if the random vector has a finite second-order moment (see \cite{ollila2012complex} for details)}. This CES distribution will be denoted by  $\mathbb{C}ES(\mub,\Mb, g_{\zb})$. In this paper, we will assume that $\mub=\0b$, as it is generally the case for many signal processing applications.

\subsubsection{Stochastic Representation Theorem} A zero mean random vector $\zb \sim \mathbb{C}ES_m(\0b,\Mb ,g_{\zb})$ if and only if it admits the following stochastic representation \cite{Yao73}
\begin{equation}\label{rep-thm}
\zb\overset{d}{=}\sqrt{Q}\Ab\ub,
\end{equation}
where the non-negative real random variable $Q$, called the modular variate, is independent of the random vector $\ub$ that is uniformly distributed on the unit complex $m$-sphere and $\Mb =\Ab\Ab^H$ is a factorization of $\Mb$.
 
\subsubsection{Circular Complex Normal Distribution}
Complex normal (Gaussian) distribution is a particular case of CES distributions in which $g_{\zb}(z)=e^{-z}$ and $C=\pi^{-m}$. Thus, the PDF of $\zb \sim  \mathbb{C}N(\0b,\Mb)$ is given by
\begin{eqnarray}
\label{pdfn}
h_{\zb}(\zb)=\frac{\exp\vg -\zb^H \Mb^{-1}\zb\vd}{\pi^m |\Mb |} .
\end{eqnarray}
Note that for $\mub=\0b$ the PDF \eqref{pdfgn} reduces to the PDF above since the scatter matrix is equal to the covariance matrix $\Cb=\Mb$ (i.e., the scale factor $\xi$ is equal to 1). Regarding the previous stochastic representation theorem, for CN-distributed vector $\zb$ the random variable $Q$ has a scaled chi-squared distribution $ Q \sim (1/2)\chi_{2m}^2$.

\subsubsection{Gaussian-core representation of CES}
In order to better explain the context of this work, we will rewrite the stochastic representation using the fact that $\ub\overset{d}{=}\nb/\|\nb\|$, where $\nb \sim \mathbb{C}N(\0b,\Iden)$. Hence, a random vector $\zb \sim \mathbb{C}ES(\0b,\Mb ,g_{\zb})$ can be represented as
\begin{equation}
\label{gaus-kern}
\zb\overset{d}{=}\frac{\sqrt{Q}}{\|\nb\|}\Ab\nb
\end{equation}
with $Q$ and $\Ab$ defined as in Eq. \eqref{rep-thm}. If $\sqrt{Q}/\|\nb\|$ is independent of $\nb$, the vector $\zb$ is said to have a compound-Gaussian distribution and it can be represented as $\zb\overset{d}{=}\sqrt{\tau}\xb$, where the non-negative real random variable $\tau$, generally called the texture, is independent of the vector $\xb\sim \mathbb{C}N(\0b,\Mb)$. 

\subsubsection{Student $t$-distribution}
A zero-mean random vector $\zb$ follows a complex multivariate $t$-distribution with $\nu$ $(0<\nu<\infty)$ degrees of freedom if the corresponding stochastic representation admits $Q \sim mF_{2m,\nu}$. This distribution belongs to the compound-Gaussian distributions where $\tau \sim \textrm{IG}(\nu/2,\nu/2)$, with $\textrm{IG}$ denoting the inverse Gamma distribution. Note that the case $\nu \to \infty$ leads to the Gaussian distribution. The multivariate $t$-distributions, besides the Gaussian distribution, encompass also multivariate Laplace distribution (for $\nu=1/2$) and the multivariate Cauchy distribution (for $\nu= 1$) which are heavy-tailed alternatives to the Gaussian distribution. The complex multivariate $t$-distributions are thus useful for studying robustness of multivariate statistics as a decrement of $\nu$ yields to distributions with an increased heaviness of the tails. We shall write $\mathbb{C}t_{\nu}(\0b,\Mb)$ to denote this case.

\subsection{Wishart distribution}
The complex Wishart distribution is the distribution of $\sum_{k=1}^{K} \xb_k\xb_k^H$, when $\xb_k$ are $m$-dimensional complex circular i.i.d. zero-mean Gaussian vectors with covariance matrix $\Mb$. Let 
\begin{equation*}
\widehat \Mb_{SCM}=\frac{1}{K}\sum_{k=1}^{K} \xb_k\xb_k^H 
\end{equation*}
be the related sample covariance matrix (SCM) which will be also referred to as a Wishart matrix. Its asymptotic distribution \cite{statistics1999}, is given by
\begin{equation}
\label{SCM}
\sqrt{K}\vec\pg\widehat \Mb_{SCM} - \Mb\pd \overset{d}{\to}  \mathbb{C}N\pg\0b,\Sig_{SCM},\Omegab_{SCM}\pd
\end{equation}
where the asymptotic covariance and pseudo-covariance matrices are
\begin{equation}
\label{asymp-SCM}
\left\{
\begin{array}{l}
\Sig_{SCM}=\Mb^{T}\otimes \Mb\\
\Omegab_{SCM}=\pg\Mb^{T}\otimes \Mb\pd\Kb.
\end{array}
\right.
\end{equation}

\subsection{$M$-estimators}

Let $(\zb_1, \hdots, \zb_K)$ be a $K$-sample of $m$-dimensional complex i.i.d. vectors with $\zb_k \sim \mathbb{C}ES(\0b,\Mb, g_{\zb}$). An $M$-estimator, denoted by $\widehat \Mb$, is defined by the solution of the following $M$-estimating equation 
\begin{equation}
\label{mest}
\widehat \Mb = \frac{1}{K}\sum_{k=1}^{K}\varphi\pg\zb_k^H \widehat \Mb^{-1}\zb_k\pd\zb_k\zb_k^H
\end{equation}
where $\varphi$ is any real-valued weight function on $[0,\infty)$ that respects Maronna's conditions\footnote{The weight function $\varphi$ does not need to be related to the PDF of any particular CES distribution, and hence $M$-estimators constitute a wide class of scatter matrix estimators.} \cite{Maronna76}.
The theoretical (population) scatter matrix $M$-functional is defined as a solution of
\begin{equation*}
\mathbb{E}\cg\varphi\pg\zb^H \Mb_{\sigma}^{-1}\zb\pd\zb\zb^H\cd =\Mb_{\sigma}
\end{equation*}
The $M$-functional is proportional to the true scatter matrix parameter $\Mb$ as
$\Mb_{\sigma}=\sigma^{-1}\Mb$, where the scalar factor $\sigma>0$ can be found by solving
\begin{equation}
\label{sig}
\mathbb{E}\cg\psi(\sigma t)\cd=m
\end{equation}
with $\psi(\sigma t)=\varphi(\sigma t)\sigma t$ and $t=\zb^H \widehat \Mb^{-1}\zb$.
\begin{Thm}
\label{thm-m}
Let $\widehat \Mb$ be a complex $M$-estimator following Maronna's conditions \cite{Maronna76}. Then
\begin{equation*}
\sqrt{K}\vec\pg\widehat \Mb- \Mb_{\sigma}\pd \overset{d}{\to} \mathbb{C}N\pg\0b,\Sig_M,\Omegab_M\pd
\end{equation*}
where the asymptotic covariance and pseudo-covariance matrices are
\begin{equation}
\label{asymp-M}
\left\{
\begin{array}{l}
\Sig_M=\vartheta_1 \Mb_{\sigma}^{T}\otimes \Mb_{\sigma}+\vartheta_2\vec( \Mb_{\sigma})\vec( \Mb_{\sigma})^{H} , \\
\Omegab_M=\vartheta_1\pg\Mb_{\sigma}^{T}\otimes \Mb_{\sigma}\pd\Kb+\vartheta_2\vec( \Mb_{\sigma})\vec( \Mb_{\sigma})^{T}.
\end{array}
\right.
\end{equation}
The constants $\vartheta_1 > 0$ and $\vartheta_2>-\vartheta_1/m$ are given in \cite{mahot2013asymptotic, ollila2012complex}.
\end{Thm}

\subsubsection{Tyler's estimator}
For the particular function $\varphi(t)=m/t$, Tyler's estimator $\widehat \Mb_{FP}$ is the solution (up to a scale factor) of the following equation
\begin{equation}
\label{FP}
\widehat \Mb_{FP}=\frac{m}{K}\sum_{k=1}^{K} \frac{\zb_k\zb_k^H}{\zb_k^H \widehat \Mb_{FP}^{-1}\zb_k}.
\end{equation}
It should be noted that for Tyler's estimator, $\Sig_T$ and $\Omegab_T$ are also defined as in Eq. \eqref{asymp-M} (see \cite{pascal2008performance}) for $\sigma=1$ and
\begin{equation}
\label{cts-FP}
\begin{array}{l}
\vartheta_1=m^{-1}(m+1)  \\
\vartheta_2=-m^{-2}(m+1)
\end{array}
\end{equation}

\subsubsection{Huber's $M$-estimator}
The complex extension of Huber's $M$-estimator is defined by
\begin{align}
\widehat{\Mb}_{Hub}=&\frac{1}{K\beta}\sum_{k=1}^{K} \cg\zb_k\zb_k^H \mathbbm{1}_{\zb_k^H \widehat \Mb_{Hub}^{-1}\zb_k\leq p^2}\cd \notag \\
&+\frac{1}{K\beta}p^2\sum_{k=1}^{K}\cg \frac{\zb_k\zb_k^H}{\zb_k^H \widehat \Mb_{Hub}^{-1}\zb_k}\mathbbm{1}_{\zb_k^H \widehat \Mb_{Hub}^{-1}\zb_k>p^2}\cd,\end{align}
where $p^2$ and $\beta$ depend on a single parameter $0<q<1$, according to $q=F_{2m}(2p^2)$ and $\beta=F_{2m+2}(2p^2)+p^2\frac{1-q}{m}$ where $F_m(\cdot)$ is the cumulative distribution function of a $\chi^2$ distribution with $m$ degrees of freedom. Note that the Huber's $M$-estimator can be interpreted as a weighted combination between the SCM and Tyler's estimator.\\

\subsubsection{Student's $M$-estimator}
The MLE for the Student $t$-distribution, denoted $\widehat \Mb_t$, is obtained as a solution of the following equation 
\begin{equation}
\label{stud}
\widehat{\Mb}_t=\frac{m+\nu/2}{K}\sum_{k=1}^{K}\frac{\zb_k\zb_k^H }{\nu/2+\zb_k^H \widehat \Mb^{-1}_t\zb_k}.
\end{equation}
The motivation to analyze this estimator arises from the fact that it presents a trade-off between the SCM and Tyler's estimator, but in a different way as for the Huber's $M$-estimator. Indeed, $\nu \to \infty$ leads to the Gaussian distribution and the resulting MLE is the SCM $(\varphi(t) \to 1)$ while $\nu\to0$ yields Tyler's estimator $(\varphi(t)\to m/t)$. Finally, $\widehat \Mb_t$ is widely used both in theory (as a benchmark) and in practice which presents strong motivation for understanding its behavior. Note also that as others $M$-estimators, it is not always used as a MLE for the $t$ distribution.

\subsection{Mahalanobis distance}
Mahalanobis distance \cite{mahalanobis1936generalized,DEMAESSCHALCK20001} is one of the most common measures in multivariate statistics and signal processing. It is based on the correlation between variables thanks to which different models can be identified and analyzed. The Mahalanobis distance of $\zb$ from $\mub$  is given by $\Delta(\mub,\Mb)$ where
\begin{equation}
\label{chi2}
\Delta^2(\mub,\Mb)=(\zb-\mub)^H\Mb^{-1}(\zb-\mub).
\end{equation}
where $\mub$ is population mean and $\Mb$ is common scatter matrix.
Since we work with zero mean vectors, we will analyze $\Delta^2(\Mb)=\zb^H\Mb^{-1}\zb$, without loss of generality. If the data are normal distributed, $\zb\sim\mathbb{C}N(\0b,\Mb)$, and the distance is based on the true scatter matrix $\Mb$, then it follows a scaled chi-squared distribution
\begin{equation}
\label{betap}
\Delta^2\pg\Mb\pd \sim \pg1/2\pd\chi^2_{2m}.
\end{equation}
Since the scatter matrix is usually unknown, the distance is computed with its estimate. If the SCM is plugged in instead of the true scatter matrix, the distance becomes $\beta'$-distributed\footnote{Beta prime distribution corresponds to a scaled F-distribution.} with an asymptotic chi-squared distribution
\begin{equation}
\Delta^2\pg\widehat{\Mb}_{SCM}\pd \sim K\beta'\pg m,K-m+1\pd
\end{equation}
where $\beta'(a,b)$ denotes a Beta prime distribution with real shape parameters $a$ and $b$. 

Beside testing if an observed random sample is from a multivariate
normal distribution (detecting outliers) \cite{rousseeuw1990unmasking,hadi1992identifying}, the Mahalanobis distance is also a useful way to determine similarities between sets of known and unknown data. Thus, it is widely used in classification problems \cite{XIANG20083600,weinberger2006distance}, feature selection problems \cite{pudil1994floating}, anomaly detection in hyperspectral images \cite{chang2002anomaly,Frontera2014}, etc. 

The object of our study is to analyze the robust Mahalanobis distance, i.e. the distance computed with $M$-estimators, comparing it to the one based on the SCM, in order to better understand its behavior. 

\section{Main contribution}
\label{sec:contrib}

This section is devoted to the main contribution of the paper. First, the results for the asymptotic distribution of the difference between any $M$-estimator and the corresponding SCM in a Gaussian context are derived. Then, the results for particular $M$-estimators and the application to Mahalanobis distance are presented. Finally, discussion and some explications are provided to emphasize the  significance of the theoretical results.

\subsection{$M$-estimators}
\label{m_est}
Basing on the previously introduced Gaussian-core model, let us assume that $K$ measurements are defined as follows:
\begin{itemize}
\item $\zb_k=\sqrt{Q_k}/\|\nb_k\|\Ab\nb_k \sim \mathbb{C}ES(\0b,\Mb)$, $k=1,\hdots,K$\\
 with
 \begin{itemize}
 \item $(\nb_1, \hdots ,\nb_K)$ be a $K$-sample of $m$-dimensional complex i.i.d. vectors with $\nb_k \sim \mathbb{C}N(\0b,\Iden)$
\item $Q_1,\hdots,Q_K$ a $K$-sample of non-negative real i.i.d random variables independent of the $\nb_k$'s
\item $\Mb=\Ab\Ab^H$ is a factorization of $\Mb$
 \end{itemize}
\end{itemize}
 $(\zb_1,\hdots,\zb_K)$ corresponds to observed data, without more specifications on their distribution, and are used to design an $M$-estimator $\widehat \Mb$. 
 
Let us also consider some ``fictive'' data (non observable) given by:
\begin{itemize}
\item $\xb_k=\Ab\nb_k \sim \mathbb{C}N(\0b,\Mb)$, $k=1,\hdots,K$
\end{itemize}
and consider the SCM $\widehat \Mb_{SCM}$ built with $(\xb_1,\hdots,\xb_K)$.
Hereafter, we always consider the same model unless it is stressed differently.

\begin{Thm}
\label{thm-1}
Let $\widehat \Mb$ be defined by Eq. \eqref{mest} and $\sigma$ is the solution of Eq. \eqref{sig}. The asymptotic distribution of $\sigma \widehat \Mb - \widehat \Mb_{SCM}$ is given by 
\begin{equation}
\label{complex_thm}
\sqrt{K}\vec\pg\sigma\widehat \Mb-\widehat \Mb_{SCM}\pd \overset{d}{\to}\mathbb{C}N \pg\0b,\Sig,\Omegab\pd
\end{equation}
where $\Sig$ and $\Omegab$ are defined by
\begin{eqnarray}
\label{result1}
\Sig&=&\sigma_1 \Mb^{T}\otimes \Mb+\sigma_2\vec( \Mb)\vec( \Mb)^{H} ,  \notag  \\
\Omegab&=&\sigma_1 \pg\Mb^{T}\otimes \Mb\pd\Kb+\sigma_2 \vec( \Mb)\vec( \Mb)^{T}
\end{eqnarray}
with 
\begin{eqnarray}
\label{sigma-complex}
\sigma_1&=&\frac{am(m+1)+c(c-2b)}{c^2} \notag \\
\sigma_2&=&\frac{a-m^2}{(c-m^2)^2}-\frac{a(m+1)}{c^2}+2\frac{m(c-b)}{c(c-m^2)}
\end{eqnarray}
where $a=E[\psi^2(\sigma t_1)]$, $b=E[\psi(\sigma t_1)t_2]$ and $c=E[\psi'(\sigma t_1)\sigma t_1]+m^2$.
\end{Thm}

\begin{remark}
\label{rem-non-null}
Notice that the structure of the asymptotic covariance matrix $\Sig$ is the same as in classical asymptotic results (Eqs.~\eqref{asymp-SCM} and \eqref{asymp-M}) but the coefficients are different. In the case of the identity matrix as covariance matrix, this very particular structure involves only three non-null elements $d_1, d_2$ and $d_3$ at the positions $(i,j)$ and equal to:
\begin{itemize}
\item $d_1=\sigma_1+\sigma_2$ for $i=j=p+m(p-1)$ with $p=1, \hdots, m$, \item $d_2=\sigma_1$ for $i=j=p+m(q-1)$ with $p\neq q$ and $p,q=1, \hdots, m$, \item $d_3=\sigma_2$ for $i=p+m(p-1),j=q+m(q-1)$ with $p\neq q$ and $p,q=1, \hdots, m$.
\end{itemize}
Similar comment with slight modifications is valid for the pseudo-covariance matrix.
\end{remark}

\begin{myproof}[Proof sketch]
We provide only a sketch of the proof, while the detailed proof of Theorem \ref{thm-1} is given in Appendix \ref{app-1}. The main idea is to represent the matrix $\Sig$ as 
\begin{equation*}
\Sig=\Sig_1(\Mb)-2\Sig_2(\Mb)+\Sig_3(\Mb)
\end{equation*}
where $\Sig_1(\Mb)$ and $\Sig_3(\Mb)$ are given by Eq. \eqref{asymp-M} and Eq. \eqref{SCM}, respectively and the matrix $\Sig_2(\Mb)$ is the correlation matrix between an $M$-estimator and the corresponding SCM in a Gaussian context. The second important step relies on a decomposition $\Sig_2(\Mb)$: 
\begin{equation*}
\Sig_2(\Mb)=\Db_1^{-1}(\Mb)\Bb_2(\Mb)
\pg\Db_2^{-1}(\Mb)\pd^H
\end{equation*}
where $\Db_1(\Mb)=E\cg d\{\vec\Psi_1(\Mb)\}/d\{\vec(\Mb\}\cd$, $\Bb_2(\Mb)= \textrm{cov}\pg\vec\Psi_1(\Mb),\vec\Psi_2(\Mb)\pd$ and $\Db_2(\Mb)=E\cg d\{\vec\Psi_2(\Mb)\}/d\{\vec(\Mb)\}\cd$ with $\Psi_1(\Mb)=\sigma \varphi(\zb^H (\sigma^{-1}\Mb)^{-1}\zb)\zb\zb^H-\Mb$ and $\Psi_2(\Mb)=\xb\xb^H-\Mb$, which is a generalization of a result derived in \cite{Maronna76}. Finally, using the dependence between the practical and fictive data one can derive elements of the matrix $\Sig_2(\Mb)$ and obtain the final result.
\end{myproof}

\begin{remark}
\label{real-case}
In this paper, we consider only complex $M$-estimators since they are used in signal processing applications. The results for the real case are given in the supplemental material. In the proof we provide only the steps that differ from the ones obtained in the complex case. It should be noted that the results of Theorem \ref{thm-1} can also be derived using the results for the real case and vector/matrix complex-to-real mapping \cite{ollila2012complex}. This is briefly discussed at the end of the additional document.
\end{remark}

\subsection{Particular cases}

\subsubsection{Tyler's estimator}
Hereafter, the results derived in \cite{draskovic2016TyE}, are presented . The first scale factor in the result can be (roughly speaking) obtained from the Theorem \ref{thm-1}\footnote{by considering the function $\psi(x)=m$.} while the derivation of the second one requires a different approach.

\begin{Thm}
\label{thm-3}
Let $\widehat \Mb_{FP}$ be defined by Eq. \eqref{FP}. The asymptotic distribution of $ \widehat \Mb_{FP} - \widehat \Mb_{SCM}$ is given by
\begin{equation*}
\sqrt{K}\vec\pg\widehat \Mb_{FP}-\widehat \Mb_{SCM}\pd \overset{d}{\to}\mathbb{C}N \pg\0b,\Sig_{FP},\Omegab_{FP}\pd
\end{equation*}
where $\Sig_{FP}$ and $\Omegab_{FP}$ are defined by

\begin{eqnarray*}
\Sig_{FP}&=&\frac{1}{m} \Mb^{T}\otimes \Mb+\frac{m-1}{m^{2}}\vec( \Mb)\vec( \Mb)^{H} \notag, \\
\Omegab_{FP}&=&\frac{1}{m}\pg\Mb^{T}\otimes \Mb\pd\Kb+\frac{m-1}{m^{2}}\vec( \Mb)\vec( \Mb)^{T}.
\end{eqnarray*}
\end{Thm}

\begin{remark}\,
In the case of Tyler's estimator $\psi(\sigma t_1 )=m$ which leads to $a=m^2$, $b=m^2$ and $c=m^2$. Substituting these values in the expression of $\sigma_1$, one obtains the previous result. This is in agreement with the results obtained in \cite{draskovic2016TyE,draskovic2017TyEANMF}.

\end{remark}

\subsubsection{Student's $M$-estimator}
In this subsection, one gives the results for the Student's $M$-estimator and $t$-distributed data. Let 
\begin{itemize}
\item $\xb_k\sim \mathbb{C}N(\0b,\Mb)$, $k=1,...,K$
\item $\tau_k\sim \textrm{IG}(\nu/2,\nu/2)$, $k=1,...,K$
\item $\zb_k=\sqrt{\tau_k}\xb_k \sim \mathbb{C}t_{\nu}(\0b,\Mb)$, $k=1,...,K$
\end{itemize}
where $\Mb=\Ab\Ab^H$ is a factorization of $\Mb$. Consider the SCM $\widehat \Mb_{SCM}$ built with $(\xb_1,\hdots,\xb_K)$ and the Student's $M$-estimator $\widehat \Mb_t$ built with $(\zb_1,\hdots,\zb_K)$.
\begin{corollary}
\label{cor}
Let $\widehat{\Mb}_t$ be defined by Eq. \eqref{stud}. The asymptotic distribution of $\widehat \Mb_t-\widehat \Mb_{SCM}$ is given by \eqref{result1} with $\sigma_1=(m+\nu/2)^{-1}$ and $\sigma_2=2/\nu(m+1+\nu/2)(m+\nu/2)^{-1}$.
\end{corollary}
\begin{IEEEproof} 
See Appendix \ref{app-2}.\\
\end{IEEEproof}

\subsubsection{Huber's $M$-estimator}
The theoretical derivation of the asymptotical distribution for Huber's $M$-estimator is impossible, since the function $\psi(t_1)$ is not differentiable in each point. However, we will present empirical results for this estimator in the next section.

\subsection{Application to Mahalanobis distance}
In this subsection we provide results for the robust Mahalanobis distance which shows the main interest of our contribution. 
\begin{Thm}
\label{distance}
Let $\widehat \Mb$ be defined by Eq. \eqref{mest} and $\sigma$ is the solution of \eqref{sig}. For the Mahalanobis distance based on $\sigma \widehat \Mb$ one has, conditionally to the distribution of $\zb$, the following asymptotic distribution
\begin{equation}
\label{mah_asm}
\sqrt{K} \pg\frac{\zb^H(\sigma\widehat \Mb)^{-1}\zb-\zb^H\widehat \Mb_{SCM}^{-1}\zb}{\zb^H \Mb^{-1}\zb}\pd _{\zb} \overset{d}{\to} N(0,\phi)
\end{equation}
where
\begin{equation}
\label{phi}
\phi=\sigma_1+\sigma_2
\end{equation}
with $\sigma_1$ and $\sigma_2$ given by Eq. \eqref{sigma-complex} and where the notation $(.)_{\zb}$ stresses the conditional distribution, conditional to $\zb$.
\end{Thm}
\begin{IEEEproof} 
See Appendix \ref{app-3}.
\end{IEEEproof}
\begin{remark}\,
The asymptotic variance of the robust Mahalanobis distance when centering around Wishart-based distance is smaller than the one when centering around the distance based on the true scatter matrix since $\sigma_1+\sigma_2<\vartheta_1+\vartheta_2$. The results are accurate even when $K$ is small which will be demonstrate in the simulation section. These findings reveal that the distribution of the robust (squared) Mahalanobis distance is better approximated with a scaled Beta prime distribution than with a scaled chi-squared distribution.
\end{remark}
\subsection{Discussion}
Here are some general comments on the proposed results as well as their great interest in practice.

\begin{enumerate}
\item First, to examine the values of the scale factors in Eq. \eqref{sigma-complex}, we discuss the values of $E[\psi^2(\sigma t_1)]$, $E[\psi(\sigma t_1)t_2]$ and $E[\psi'(\sigma t_1)\sigma t_1]+m^2$. Since $0<\psi(\sigma t_1)<M$ and $E(\psi(\sigma t_1))=m$, using Bhatia-Davis inequality \cite{bhatia2000better}, one has that $\textrm{var}(\psi(\sigma t_1))<(M-m)m$ and thus $ E(\psi(\sigma t_1)^2)<Mm$. Since $M$ is of same magnitude as $m$ and $M>m$, one obtains that $ E(\psi(\sigma t_1)^2)$ is of  same magnitude as $m^2$ (for Tyler's estimator $E(\psi(\sigma t_1)^2)=m^2$, for Student $M$-estimator $ E(\psi(\sigma t_1)^2)= \frac{m(m+1)(m+\frac{\nu}{2})}{m+1+\frac{\nu}{2}}$, for SCM $ E(\psi(\sigma t_1)^2)=m^2+m$...). From this, it follows that $b$ is also of the same magnitude as $m^2$ since $m^2= E(\psi(\sigma t_1))E(t_2)<E(\psi(\sigma t_1)t_2)<\sqrt{ E(\psi(\sigma t_1)^2)}\sqrt{E(t_2^2)}< \sqrt{(m^2+m)}\sqrt{ E(\psi(\sigma t_1)^2)}$. It is obvious that $c$ is also of the same magnitude as $m^2$. Generally, for all widely used $M$-estimators, one obtains that $a,b,c=m^2+\alpha m$, $\alpha>0$ which leads to $\sigma_1$ inversely proportional to $m$. For $\sigma_2$, one can not provide precise information about its value, but it turns out that it is eather smaller (e.g., Tyler's estimator) or unchanged (e.g., Student's $M$-estimator) comparing to the scale factor given in Eq. \eqref{asymp-M}. This ensures the strong ``proximity'' between $M$-estimators and SCM, justifying the approximation of $M$-estimators behavior thanks to a Wishart distribution.\\

\item The results derived in this paper show that all $M$-estimators are asymptotically closer to the SCM than to the true covariance matrix. By ``close'', we mean that the asymptotic variance when centering about the Gaussian-based SCM is much smaller than the one when centering about true scatter matrix. Also, this difference is more obvious when the dimension $m$ increases. This remark is of course also obvious for Tyler's estimator. \\

\item An important consequence of the previous remark is that any $M$-estimators (including Tyler's one) behavior can be approximated by the SCM one (built with Gaussian random vectors), namely by the Wishart distribution. This is of great interest in practice since all the analytical performance of functionals of robust scatter estimators can be derived based on its equivalent for the simplest Wishart distribution, while keeping the inherent robustness brought by $M$-estimators (contrary to the SCM). To summarize, robust estimators are better approximated by Wishart distribution than by the asymptotic Gaussian distribution with the true scatter matrix as mean.\\

\item Another comment is that, roughly speaking, one has the following result for any robust scatter matrix estimator $\hat{\Mb}$:
$$\sqrt{mK}\pg\hat{\Mb}-\hat{\Mb}_{SCM}\pd \xrightarrow[K\to \infty]{} \mathbb{C}N \pg\0b,\Db,\mathbf Q\pd$$
where $\Db$ and $\mathbf Q$ are ``fixed''. Thus, one has a gain in terms of convergence of $m$. This is agreement with the results obtained in \cite{couillet2015therandom} for a different convergence regime ($m,K \to \infty$ with $m/K$ tending to a positive constant).\\

\item Finally, it should be pointed out that the results can be applied to various signal processing problems. One can note that the scaled variance of the robust Mahalanobis distance when centering around the one based on the SCM in a Gaussian context depends only on the scale factors given by Eq. \eqref{sigma-complex}. This directly leads to the conclusion that the distribution of the robust distance can be better approximated with the one of the SCM-based distance, than with the asymptotical chi-squared distribution. These results can be  extended to various problems such as detection or classification problems (see e.g., \cite{draskovic2017TyEANMF}).
\end{enumerate}

\section{Simulations}
\label{sec:simus}
\subsection{Validation of the theoretical results}

In this section we first present some simulations that validate the theoretical results of Theorem \ref{thm-1}.  Figure \ref{fig1} presents the empirical mean\footnote{obtained as the empirical mean of the quantities obtained from $I$ Monte Carlo runs ($I =10K$)} norm of the difference between the empirical covariance matrix of $\sqrt{K}(\sigma\widehat \Mb-\widehat \Mb_{SCM})$ (Eq. \eqref{complex_thm}), denoted as $\Sig^{(K)}$ and the theoretical results obtained in Theorem \ref{thm-1}. The plotted results are obtained from $t$-distributed data with a DoF $\nu$ set to 2 and using the Student's $M$-estimator (for which theoretical results are explicitly given in Corollary \ref{cor}). \begin{figure}[h!]
\begin{center}
{\label{err-cv=0.99}\begin{tikzpicture}[font=\footnotesize,scale=0.9]
\pgfplotsset{every axis/.append style={mark options=solid, mark size=2.5pt}}
\pgfplotsset{every axis legend/.append style={fill=white,cells={anchor=west},at={(0.50,0.98)},anchor=north west}} \tikzstyle{every axis y label}+=[yshift=-10pt]
\tikzstyle{every axis x label}+=[yshift=5pt]
\tikzstyle{dashed dotted}=[dash pattern=on 1pt off 4pt on 6pt off 4pt]

\begin{axis}[xlabel={$K$},ylabel={$\| \Cb^{(K)}-\Cb\|$}, xmode=log, ymode=log]

\addplot[smooth,blue,line width=.5pt] plot coordinates {
(10.000000,1.392378)(12.000000,0.861929)(14.000000,0.745893)(17.000000,0.471459)(20.000000,0.472314)(24.000000,0.389181)(29.000000,0.319734)(35.000000,0.250467)(41.000000,0.244146)(49.000000,0.140681)(59.000000,0.099867)(70.000000,0.090589)(84.000000,0.113766)(100.000000,0.079077)(119.000000,0.040022)(143.000000,0.062990)(170.000000,0.057036)(203.000000,0.052406)(242.000000,0.023621)(289.000000,0.033143)(346.000000,0.025068)(412.000000,0.016062)(492.000000,0.013545)(588.000000,0.017782)(702.000000,0.013396)(838.000000,0.007568)(1000.000000,0.007157)(1194.000000,0.010301)(1425.000000,0.004293)(1701.000000,0.002978)(2031.000000,0.004232)(2424.000000,0.002784)(2894.000000,0.003402)(3455.000000,0.001888)(4125.000000,0.002103)(4924.000000,0.003074)

};

\end{axis}
\end{tikzpicture}
}

\vspace*{-0.3cm}
\caption{\small \label{fig1}  Euclidean norm of the difference between the empirical  covariance matrix of Eq. \eqref{complex_thm} and the theoretical result Eq. \eqref{result1} with $\sigma_1$ and $\sigma_2$ of Corollary \ref{cor}. $m=5$}
\end{center}
\end{figure}
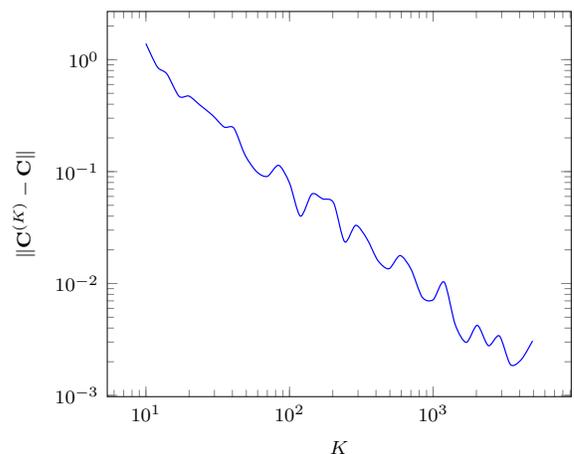
 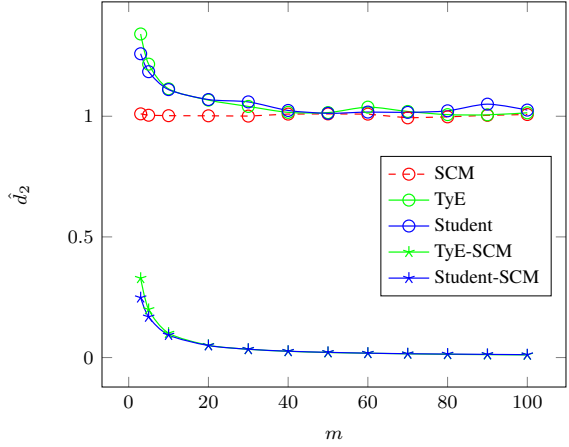
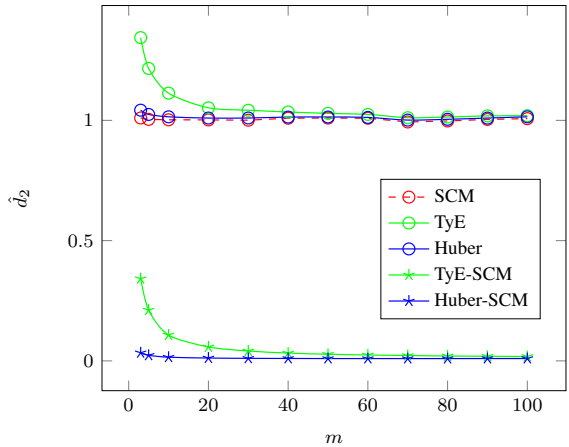
\begin{figure}[!h]
\captionsetup{singlelinecheck=off}
\begin{center}

\subfigure[Student-t distribution]{\label{t-dist}\begin{tikzpicture}[font=\footnotesize,scale=0.9]
\pgfplotsset{every axis/.append style={mark options=solid, mark size=2.5pt}}
\pgfplotsset{every axis legend/.append style={fill=white,cells={anchor=west},at={(0.60,0.60)},anchor=north west}} \tikzstyle{every axis y label}+=[yshift=-10pt]
\tikzstyle{every axis x label}+=[yshift=5pt]
\tikzstyle{dashed dotted}=[dash pattern=on 1pt off 4pt on 6pt off 4pt]

\begin{axis}[xlabel={$m$},ylabel={$\hat d_2$}]
\addplot[mark=o,dashed,red,line width=.5pt] plot coordinates {
(3.000000,1.009876)(5.000000,1.004123)(10.000000,1.002357)(20.000000,1.001839)(30.000000,1.000674)(40.000000,1.008968)(50.000000,1.009699)(60.000000,1.008821)(70.000000,0.993370)(80.000000,0.997491)(90.000000,1.003794)(100.000000,1.007189)
};
\addplot[mark=o,smooth,green,line width=.5pt] plot coordinates {
(3.000000,1.340950)(5.000000,1.215883)(10.000000,1.113570)(20.000000,1.066947)(30.000000,1.040698)(40.000000,1.016187)(50.000000,1.015063)(60.000000,1.037760)(70.000000,1.019784)(80.000000,1.006712)(90.000000,1.005790)(100.000000,1.014235)
};
\addplot[mark=o,smooth,blue,line width=.5pt] plot coordinates {
(3.000000,1.258997)(5.000000,1.184658)(10.000000,1.110376)(20.000000,1.069232)(30.000000,1.059512)(40.000000,1.024013)(50.000000,1.012192)(60.000000,1.017281)(70.000000,1.015992)(80.000000,1.022533)(90.000000,1.050124)(100.000000,1.025577)
};
\addplot[mark=star,smooth,green,line width=.5pt] plot coordinates {
(3.000000,0.328836)(5.000000,0.199518)(10.000000,0.101434)(20.000000,0.050965)(30.000000,0.034355)(40.000000,0.025928)(50.000000,0.020886)(60.000000,0.018026)(70.000000,0.015461)(80.000000,0.013393)(90.000000,0.012068)(100.000000,0.010920)
};
\addplot[mark=star,smooth,blue,line width=.5pt] plot coordinates {
(3.000000,0.248259)(5.000000,0.168695)(10.000000,0.093474)(20.000000,0.049883)(30.000000,0.034490)(40.000000,0.026311)(50.000000,0.021701)(60.000000,0.018254)(70.000000,0.015915)(80.000000,0.014521)(90.000000,0.013186)(100.000000,0.011940)
};

\legend{{SCM},{TyE},{Student},{TyE-SCM},{Student-SCM}};
\end{axis}
\end{tikzpicture}
}

\subfigure[Gaussian plus outliers]{\label{outlier}\begin{tikzpicture}[font=\footnotesize,scale=0.9]
\pgfplotsset{every axis/.append style={mark options=solid, mark size=2.5pt}}
\pgfplotsset{every axis legend/.append style={fill=white,cells={anchor=west},at={(0.60,0.55)},anchor=north west}} \tikzstyle{every axis y label}+=[yshift=-10pt]
\tikzstyle{every axis x label}+=[yshift=5pt]
\tikzstyle{dashed dotted}=[dash pattern=on 1pt off 4pt on 6pt off 4pt]

\begin{axis}[xlabel={$m$},ylabel={$\hat d_2$}]
\addplot[mark=o,dashed,red,line width=.5pt] plot coordinates {
(3.000000,1.009876)(5.000000,1.004123)(10.000000,1.002357)(20.000000,1.001839)(30.000000,1.000674)(40.000000,1.008968)(50.000000,1.009699)(60.000000,1.008821)(70.000000,0.993370)(80.000000,0.997491)(90.000000,1.003794)(100.000000,1.007189)
};
\addplot[mark=o,smooth,green,line width=.5pt] plot coordinates {
(3.000000,1.343937)(5.000000,1.215823)(10.000000,1.113151)(20.000000,1.052097)(30.000000,1.042085)(40.000000,1.034861)(50.000000,1.029004)(60.000000,1.024687)(70.000000,1.010994)(80.000000,1.013792)(90.000000,1.018515)(100.000000,1.019504)
};
\addplot[mark=o,smooth,blue,line width=.5pt] plot coordinates {
(3.000000,1.042308)(5.000000,1.024601)(10.000000,1.014680)(20.000000,1.009534)(30.000000,1.009757)(40.000000,1.013671)(50.000000,1.013965)(60.000000,1.012252)(70.000000,1.001176)(80.000000,1.004399)(90.000000,1.009495)(100.000000,1.014685)
};
\addplot[mark=star,smooth,green,line width=.5pt] plot coordinates {
(3.000000,0.341817)(5.000000,0.211047)(10.000000,0.107548)(20.000000,0.058034)(30.000000,0.041154)(40.000000,0.032541)(50.000000,0.027633)(60.000000,0.024527)(70.000000,0.022381)(80.000000,0.020352)(90.000000,0.019243)(100.000000,0.018159)
};
\addplot[mark=star,smooth,blue,line width=.5pt] plot coordinates {
(3.000000,0.033661)(5.000000,0.023153)(10.000000,0.015334)(20.000000,0.011852)(30.000000,0.010579)(40.000000,0.010118)(50.000000,0.009686)(60.000000,0.009646)(70.000000,0.009581)(80.000000,0.009583)(90.000000,0.009736)(100.000000,0.009750)
};

\legend{{SCM},{TyE},{Huber},{TyE-SCM},{Huber-SCM}};
\end{axis}
\end{tikzpicture}
}

\captionsetup{singlelinecheck=off}
\caption{\small \label{fig2-fred} Empirical second diagonal element in the asymptotic covariance matrices versus the dimension $m$ for $K=1000$. \begin{itemize} \item SCM corresponds to the empirical version of $1$ in Eq.~\eqref{asymp-SCM} \item TyE: empirical version of $\vartheta_1=m^{-1}(m+1)$ in Eq.~\eqref{cts-FP}, \item Huber: empirical version of $\vartheta_1$ in Eq.~\eqref{asymp-M} for Huber $M$-estimator, \item Student: empirical version of  $\vartheta_1$ in Eq.~\eqref{asymp-M} for the Student $M$-estimator, \item TyE-SCM: empirical version of $1/m$ of Theorem \ref{thm-3}, \item Huber-SCM: empirical version of $\sigma_1$ of Theorem \ref{thm-1} computed with Huber $M$-estimator, \item Student-SCM: empirical version of $\sigma_1=(m+\nu/2)^{-1}$ of Corollary \ref{cor}\end{itemize}}
\end{center}
\end{figure}
The scatter matrix $\Mb$ is defined by $M_{i,j}=\rho^{|i-j|}, \quad i=1,...,m.$ The correlation coefficient $\rho$ is set to 0, i.e. the scatter matrix is equal to the identity matrix. One can notice that Figure \ref{fig1} validates results obtained in Theorem \ref{thm-1} since the quantity tends to zero when the number $K$ of samples tends to infinity. 

Recall that following Remark \ref{rem-non-null} when the scatter matrix is equal to identity, the matrices $\Sig$ and $\Omegab$ contain only three different non-null elements: $\sigma_1+\sigma_2$, $\sigma_1$ and $\sigma_2$.
Here, we will compare the empirical value of $\sigma_1$ to the empirical value of $\vartheta_1$ (first scale factor of the empirical  covariance matrix of $\sqrt{K}(\sigma\widehat \Mb- \Mb)$) in distinct non-Gaussian environments. Results are similar for other coefficients and will be omitted.  

Figure \ref{fig2-fred} presents results in various non-Gaussian cases. On Figure \ref{t-dist}, the results obtained for complex $t$-distributed data ($\nu=2$) are presented. The second diagonal element for Tyler's and Student's $M$-estimator are plotted. The horizontal scale presents the dimension of the data. The number of samples $K$ is set to 1000. One can notice that the second diagonal element for $M$-estimators vanishes when $m$ increases, as expected. Indeed, if we look at the results from Theorem \ref{thm-3} and Corollary \ref{cor}, the first scale factor is inversely proportional to the dimension $m$.  

On Figure \ref{outlier}, we present the results for Tyler's and Huber's $M$-estimators when the data are corrupted by some outliers. The parameter $q$ for Huber's $M$-estimator is set to $0.95$, which means that $95\%$ of the data are considered to be Gaussian distributed while the remaining $5\%$ are treated as outliers. As it can be noted, the results are the same as on Figure \ref{t-dist}, showing the robustness of these two estimators and validating the theoretical results. Generally speaking, these tests show that $M$-estimators are better characterized by a Wishart distribution than a Gaussian distribution centered on the true matrix $\Mb$.

\subsection{Application to Mahalanobis distance}

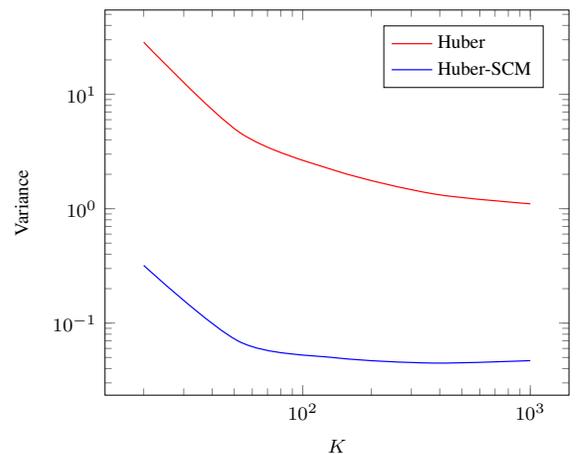
\begin{figure}[!h!]
\begin{center}

%
%

\subfigure[Tyler's estimator]{\label{tyler}\begin{tikzpicture}[font=\footnotesize,scale=0.9]
\pgfplotsset{every axis/.append style={mark options=solid, mark size=2.5pt}}
\pgfplotsset{every axis legend/.append style={fill=white,cells={anchor=west},at={(0.60,0.96)},anchor=north west}} \tikzstyle{every axis y label}+=[yshift=-10pt]
\tikzstyle{every axis x label}+=[yshift=5pt]
\tikzstyle{dashed dotted}=[dash pattern=on 1pt off 4pt on 6pt off 4pt]

\begin{axis}[xlabel={$K$},ylabel={Variance}, xmode=log, ymode=log]
\addplot[smooth,red,line width=.5pt] plot coordinates {
(20.000000,47.517593)(53.000000,5.709049)(141.000000,2.160424)(376.000000,1.398279)(1000.000000,1.122182)
};
\addplot[smooth,blue,line width=.5pt] plot coordinates {
(20.000000,3.111384)(53.000000,0.363368)(141.000000,0.237399)(376.000000,0.207941)(1000.000000,0.196273)
};
\addplot[dashed,green,line width=.5pt] plot coordinates {
(20.000000,0.190000)(53.000000,0.190000)(141.000000,0.190000)(376.000000,0.190000)(1000.000000,0.190000)
};

\legend{{TyE},{TyE-SCM},{Eq. \eqref{phi}}};
\end{axis}
\end{tikzpicture}
}

\subfigure[Huber's $M$-estimator]{\label{huber}\begin{tikzpicture}[font=\footnotesize,scale=0.9]
\pgfplotsset{every axis/.append style={mark options=solid, mark size=2.5pt}}
\pgfplotsset{every axis legend/.append style={fill=white,cells={anchor=west},at={(0.60,0.96)},anchor=north west}} \tikzstyle{every axis y label}+=[yshift=-10pt]
\tikzstyle{every axis x label}+=[yshift=5pt]
\tikzstyle{dashed dotted}=[dash pattern=on 1pt off 4pt on 6pt off 4pt]

\begin{axis}[xlabel={$K$},ylabel={Variance}, xmode=log, ymode=log]

\addplot[smooth,red,line width=.5pt] plot coordinates {
(20.000000,28.595501)(53.000000,4.617415)(141.000000,2.134958)(376.000000,1.346228)(1000.000000,1.103572)
};
\addplot[smooth,blue,line width=.5pt] plot coordinates {
(20.000000,0.319648)(53.000000,0.068415)(141.000000,0.049615)(376.000000,0.044711)(1000.000000,0.046980)
};

\legend{{Huber},{Huber-SCM}};
\end{axis}
\end{tikzpicture}
}

\caption{\small \label{fig_dist} Scaled empirical variance of the robust Mahalanobis distance compared to the one when centering around SCM-based distance (with the result of Theorem \ref{distance}) versus $K$, $m=10$}  

\end{center}
\end{figure}

We now present results for the robust Mahalanobis distance. On Figure \ref{tyler}, the results for Tyler's $M$-estimator are presented when data follow a complex $t$-distribution with $\nu=2$. 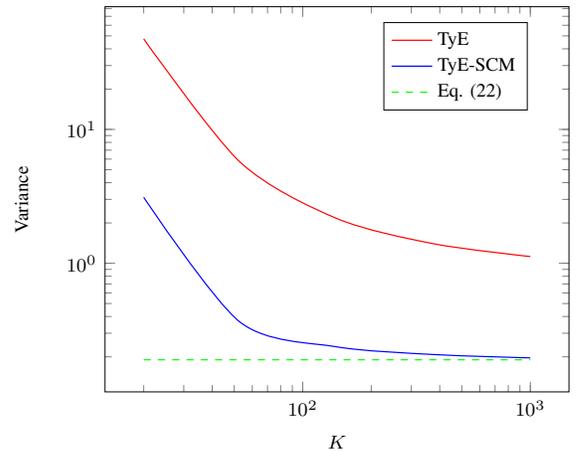

The empirical variance of the robust distance and the one of the difference between the robust distance and the distance computed with the SCM in a Gaussian context (compared to the theoretical result of Theorem \ref{mah_asm} - Eq. \eqref{phi}) are plotted.  On Figure \ref{huber} the results for Huber's $M$-estimator are plotted. $95\%$ of the data follow a Gaussian distribution, while the outliers (remaining $5\%$ of the data) are modelled with $t$-distribution ($\nu=2$). One can notice that the value of the robust distance is much closer to the one based on the SCM, than to the distance computed with the true scatter matrix, which once again justifies the statement that the behavior of $M$-estimators can be approximated by a Wishart distribution. This also implies that the distribution of robust distances can be better approximated with a theoretical distribution of the SCM-based distance in the Gaussian framework than with the asymptotic  distribution based on the true scatter matrix.
\begin{figure}[!h]
\begin{center}
\includegraphics[scale=0.55]{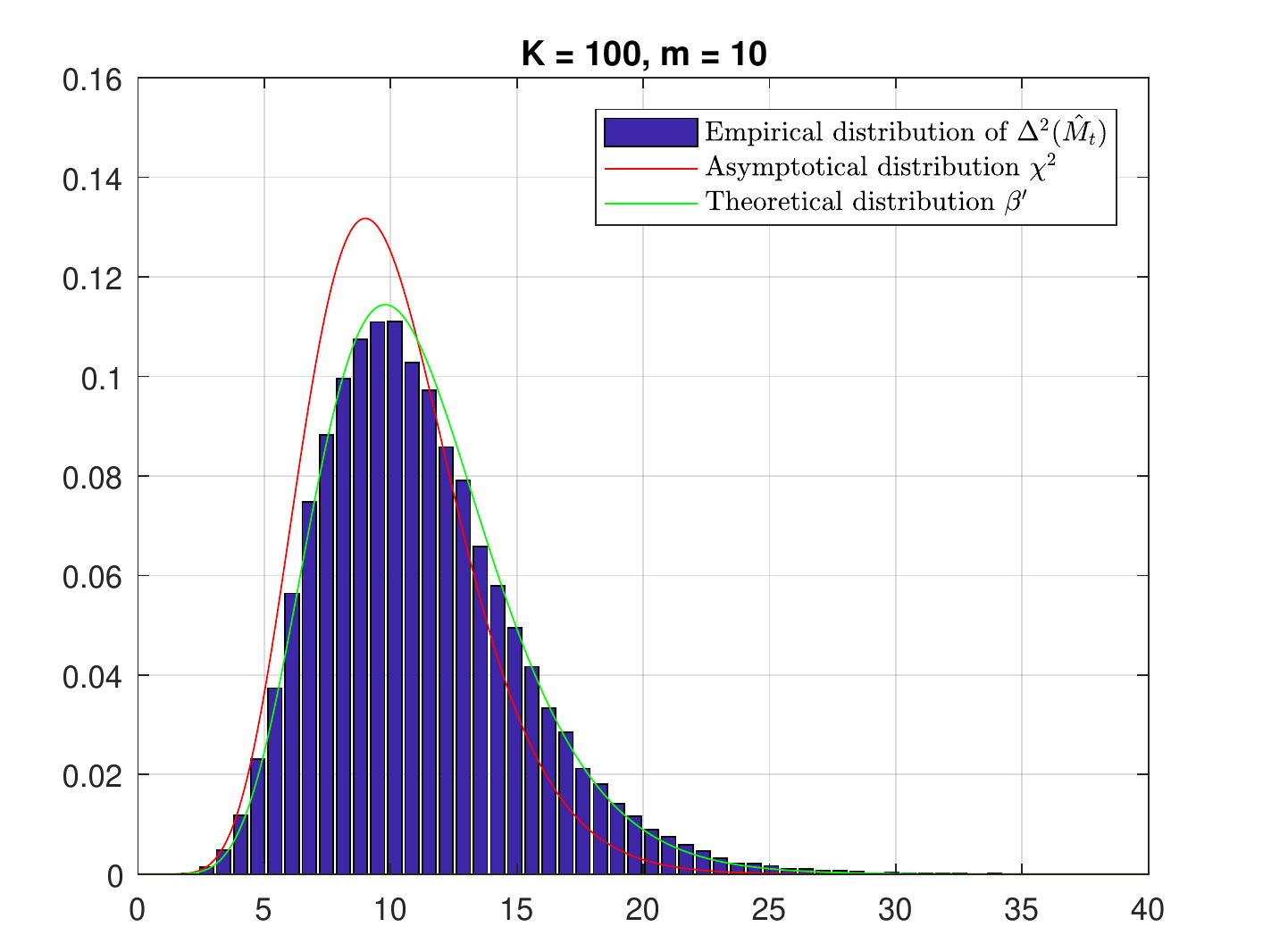}
\caption{\label{mah_distr} Histogram distribution of the robust Mahalanobis distance based on Student's $M$-estimator $\Delta^2(\widehat{\Mb}_t)$ versus asymptotic distribution (Eq. \eqref{chi2}) in red and versus theoretical approximative distibution (Eq. \eqref{betap}) in green where $m = 10$, $K = 100$, $\zb\sim\mathbb{C}t_{\nu}$ with $\nu=2$.}
\end{center}
\end{figure}

Figure \ref{mah_distr} presents the empirical distribution of the robust Mahalanobis distance built with the Student's $M$-estimator and two corresponding distributions proposed in Eqs. \eqref{chi2} and \eqref{betap} for $t$-distributed data with $\nu=2$ $(K=100, m = 10)$. We observe that the empirical distribution matches significantly better the scaled Beta prime than the scaled chi-squared distribution. The essential advantage of these findings in, for instance, outliers detection is that they support the idea to use the robust $M$-estimators to estimate the scatter matrix and to rely on the theoretical distribution of the Wishart-based distance when computing the detection treshold.

\section{Conclusions}
\label{sec:conclu}
This paper investigated the statistical properties of $M$-estimators. To that end, a new ``Gaussian-core'' model has been introduced for CES distributions. We have proposed a new approach that consists in comparing $M$-estimators to the well-known Gaussian-based SCM in order to derive new properties. In other words, the approach can be summarized as follows: explaining the behavior of an ``intractable'' estimator $\hat{\theta}$ by analyzing its proximity with a well-known estimator $\hat{\theta_1}$. It has been shown that the second order statistics of $M$-estimators when centering around a Wishart distributed matrix are much smaller than the ones when centering around the true scatter matrix. It has also been revealed that this difference is even more meaningful for high-dimensional data. It should be stressed that these results provide a better approximation of $M$-estimators properties than any other analyses in the literature. In our view, these results represent an excellent initial step toward the better understanding of the behavior of $M$-estimators applied in various problems. In this work, we have presented the application to the widely used Mahalanobis distance. This approach could also be applied to the adaptive detection problems and thus, be very helpful in the improvement of the detection performances. Moreover, one potential application of our findings can be found in polarimetric SAR images restoration, clustering and/or target detection. To conclude, we are confident that the results of this work are very promising and that can be applied to a wide range of signal processing problems. 

\appendices
\section{Proof of Theorem \ref{thm-1}}
\label{app-1}

To prove the statement let us rewrite the right hand side of equation (\ref{result1}) as follows:
\begin{eqnarray*}
&&\sqrt{K}\pg\vec(\sigma\widehat \Mb-\widehat \Mb_{SCM})\pd= \notag \\
&&\sqrt{K}\pg\vec(\sigma\widehat \Mb- \Mb -\widehat \Mb_{SCM} + \Mb )\pd \notag \\
&&=\begin{bmatrix} \mathbf{1},  \mathbf{-1}  \end{bmatrix} \begin{bmatrix} \sqrt{K} \vec(\sigma\widehat \Mb- \Mb) \\
 \sqrt{K} \vec(\widehat \Mb_{SCM} - \Mb ) \end{bmatrix}
\end{eqnarray*}
Therefore one has $\Sig^{(K)}=\Sig_1^{(K)}-2 \Sig_2^{(K)}+ \Sig_3^{(K)}$
with
\begin{eqnarray*}
\Sig_1^{(K)}&=&KE\cg\vec(\sigma\widehat \Mb-\Mb) \vec(\sigma\widehat \Mb-\Mb) ^H\cd  \\
\Sig_2^{(K)}&=&KE\cg\vec(\sigma\widehat \Mb- \Mb)  \vec(\widehat \Mb_{SCM}-\Mb )^H\cd   \\
\Sig_3^{(K)}&=&KE\cg\vec(\widehat \Mb_{SCM}- \Mb )\vec(\widehat \Mb_{SCM}- \Mb )^H\cd. 
\end{eqnarray*}
One has now 
\begin{equation}
\label{sig-final}
\Sig^{(K)}\xrightarrow [K\to+\infty]{} \Sig=\Sig_1(\Mb)-2 \Sig_2(\Mb)+ \Sig_3(\Mb)
\end{equation} 
where the matrices $\Sig_1(\Mb)$ and $\Sig_3(\Mb)$ are given by \eqref{asymp-M} and \eqref{asymp-SCM}, respectively. 

Following the similar ideas used in \cite{Maronna76,tyler1987distribution}, we provide a more general result that allows to compute a corelation between two estimators
\begin{equation*}
\Sig_2^{(K)}\xrightarrow [K\to+\infty]{}\Sig_2(\Mb)=\Db_1^{-1}(\Mb)\Bb(\Mb)
\Db_2^{-1}(\Mb) 
\end{equation*}
where $\Db_1(\Mb)=E\cg d\{\vec\Psi_1(\Mb)\}/d\{\vec(\Mb\}\cd$, $\Bb(\Mb)= \textrm{cov}\pg\vec\Psi_1(\Mb),\vec\Psi_2(\Mb)\pd$ and $\Db_2(\Mb)=E\cg d\{\vec\Psi_2(\Mb)\}/d\{\vec(\Mb)\}\cd$ with $\Psi_1(\Mb)=\sigma \varphi(\zb^H (\sigma^{-1}\Mb)^{-1}\zb)\zb\zb^H-\Mb$ and $\Psi_2(\Mb)=\xb\xb^H-\Mb$.

Without loss of generality, we will assume that $\Mb=\Iden$. Indeed, one has that 
\begin{eqnarray*}
\Sig_2(\Mb)=\pg\Mb^{T/2} \otimes \Mb^{1/2}\pd\Sig_2 (\Iden) \pg\Mb^{T/2} \otimes \Mb^{1/2}\pd^H. 
\end{eqnarray*}

In order to determine the final result, we will derive the expression for $\Sig_2(\Iden)$. One can show that 
\begin{equation*}
\Db_1^{-1}(\Iden)=\alpha_1\Iden+\alpha_2\vec(\Iden)\vec(\Iden)^T
\end{equation*}
where $\alpha_1=-\frac{m(m+1)}{c}$ and $\alpha_2=\frac{m(c^2-m^2-m)}{c(c-m^2)}$ with $c=E\cg\sigma t_1\psi'(\sigma t_1)\cd+m^2$.  Moreover, it is simple to show that $\Db_2(\Iden)^{-1}=-\Iden$. 

Then, basing on Theorem 2 from \cite{tyler1987distribution}, one can derive more general result
\begin{eqnarray*}
\Bb_2(\Iden)=\beta_1\Iden+\beta_2\vec(\Iden)\vec(\Iden)^T
\end{eqnarray*}
where 
\begin{eqnarray*}
\beta_1&=&\textrm{cov}[{\Psi_1( \Iden)}_{jk}{\Psi_2(\Iden)}_{jk}] \notag \\
&=&E[\varphi(\sigma  t_1) t_2 u_j^2 u_k^2] \notag \\
&=&\frac{E[\psi(\sigma  t_1) t_2]}{m(m+1)}=\frac{b}{m(m+1)}
\end{eqnarray*}
and 
\begin{eqnarray*}
\beta_2&=&\textrm{cov}[{\Psi_1( \Iden)}_{jj}{\Psi_2(\Iden)}_{kk}] \notag \\
&=&\beta_1-E[\psi(\sigma t_1)]E[ t_2]/m^2 \notag \\
&=&\beta_1-1
\end{eqnarray*}
 since $u_i^2 \sim Beta(1, m-1)$, $E[u_j^2]=1/m$ and $E[u_j^2u_k^2]=1/(m(m+1))$.
After some mathematical manipulations, one obtains
\begin{equation*}
\Sig_2(\Iden)=\gamma_1\Iden+\gamma_2\vec(\Iden)\vec(\Iden)^T
\end{equation*} 
with
\begin{eqnarray}
\label{gamma}
\gamma_1&=&-\alpha_1\beta_1=\frac{b}{c} \notag \\
\gamma_2&=&-(\alpha_1\beta_2+2\alpha_2\beta_1+2m\alpha_2\beta_2=\frac{m(b-c)}{c(c-m^2)}.
\end{eqnarray}
This leads to the final expression of $\Sig_2(\Mb)$:
\begin{equation}
\label{sig2-M}
\Sig_2(\Mb) = \gamma_1\Mb^{T}\otimes \Mb+\gamma_2\vec( \Mb)\vec( \Mb)^H
\end{equation}

Combining Eq.~\eqref{sig2-M} together with Eqs.~\eqref{asymp-M} and \eqref{asymp-SCM} in Eq.~\eqref{sig-final}, one obtains the coefficients $\sigma_1$ and $\sigma_2$ as follows
\begin{eqnarray*}
\sigma_1&=&\vartheta_1-2\gamma_1+1 \notag \\
&=&\frac{am(m+2)+c(c-2b)}{c^2} 
\end{eqnarray*}
and
\begin{eqnarray*}
\sigma_2&=&\vartheta_2-2\gamma_2 \notag \\
&=&\frac{a-m^2}{(c-m^2)^2}-\frac{a(m+2)}{c^2}+2\frac{m(c-b)}{c(c-m^2)}. \notag \\
\end{eqnarray*}
Finally, one can easily prove that $\Omegab=\Sig\Kb$ \cite{mahot2013asymptotic}, which leads to the final results and concludes the proof.

\section{Proof of Corollary \ref{cor}}
\label{app-2}
For Student's $t$-distribution $t_1 \sim mF_{2m,\nu}$  yields 
\begin{eqnarray*}
f(t_1)=C_mt_1^{m-1}\pg 1+\frac{2}{\nu}t_1\pd^{-\frac{2m+\nu}{2}}
\end{eqnarray*}
with
\begin{eqnarray*}
C_m=\pg\frac{2}{\nu}\pd ^m\frac{\Gamma\pg m+\frac{\nu}{2}\pd}{\Gamma\pg m\pd\Gamma\pg\frac{\nu}{2}\pd}
\end{eqnarray*}
where $\Gamma(.)$ is the Gamma function.
Since $\sigma=1$ for every $ML$-estimator \cite{ollila2012complex}, one has
\begin{eqnarray*}
\psi\pg\sigma t_1\pd=\psi\pg t_1\pd=\frac{2m+\nu}{\nu+2t_1}t_1
\end{eqnarray*}
Now one obtains
\begin{eqnarray*}
&&E\cg\psi^2(t_1)\cd=E\cg\frac{\pg 2m+\nu\pd ^2}{\pg\nu+2t_1\pd ^2}t_1^2\cd \notag \\ 
&=&C_m \int_{0}^{+\infty} \frac{(2m+\nu)^2t_1^2}{\nu^2(1+\frac{2}{\nu}t_1)^2}t^{m-1}\pg 1+\frac{2t_1}{\nu}\pd^{-\frac{2m+\nu}{2}}dt_1 \notag \\
&=&C_m  \frac{(2m+\nu)^2}{\nu^2} \frac{1}{C_{m+2}} \notag \\
 &=&\frac{m(m+1)(m+\frac{\nu}{2})}{m+1+\frac{\nu}{2}}
\end{eqnarray*}
and
\begin{eqnarray*}
&&E\cg t_1\psi'(t_1)\cd=E\cg\frac{(2m+\nu)\nu}{(\nu+2 t_1)^2}t_1\cd \notag \\
&=&C_m \int_{0}^{+\infty} \frac{(2m+\nu)\nu t_1}{\nu^2(1+\frac{2}{\nu}t_1)^2}t_1^{m-1}\pg1+\frac{2t_1}{\nu}\pd^{-\frac{2m+\nu}{2}}dt_1 \notag \\
&=&C_m  \frac{\nu(2m+\nu)}{\nu^2} \int_{0}^{+\infty}t_1^{-1} t_1^{m+1}\pg 1+\frac{2t_1}{\nu}\pd^{-\frac{2m+4+\nu}{2}}dt_1 \notag \\
 &=&C_m  \frac{2m+\nu}{\nu}\frac{1}{C_{m+2}} E\cg t_1^{-1}\cd 
\end{eqnarray*}
where now $t/(m+2) \sim F_{2m+4,\nu}$ or equivalently  $(m+2)/t \sim F_{\nu,2m+4}$ which gives
\begin{eqnarray*}
E\cg t_1^{-1}\cd =\frac{1}{m+2}\frac{2m+4}{2m+4-2}=\frac{1}{m+1}
\end{eqnarray*}
and finally
\begin{eqnarray*}
E\cg t\psi'(t_1)\cd=\frac{\nu}{2}\frac{m}{(m+1+\frac{\nu}{2})}.
\end{eqnarray*}
To compute $E[\psi(t_1)t_2]$ let us remind that $t_1=\tau t_2$ where $\tau$ and $t_2$ are independent,  $\tau \sim \textrm{IG}(\nu/2,\nu/2)$ and $t_2 \sim (1/2){\chi}_{2m}^2$. Thus, one can write
\begin{eqnarray*}
I=E\cg\psi(t_1)t_2\cd=C\iint_{\mathbb R_{+}^2}\frac{1}{1+\frac{2\tau}{\nu} t_2} \tau^{-\frac{\nu}{2}} \mathrm{e}^{-\frac{\nu}{2\tau}}t_2^{m+1}\mathrm{e}^{-t_2}d\tau dt_2
\end{eqnarray*}
where $C=(\frac{2m}{\nu}+1)\frac{\nu}{2}^{\frac{\nu}{2}})/(\Gamma(\frac{\nu}{2})\Gamma(m))$. The change of variable $u=\frac{2\tau}{\nu}t_2$ gives $du=\frac{2\tau}{\nu}dt_2$ and hence
\begin{eqnarray*}
I=C\int_0^\infty \frac{\nu}{2\tau}\tau^{-\frac{\nu}{2}}\mathrm{e}^{-\frac{\nu}{2\tau}}\big(\frac{\nu}{2\tau}\big)^{m+1}\int_0^\infty \frac{1}{1+u} u^{m+1}\mathrm{e}^{-\frac{\nu}{2\tau}u}du d\tau \,.
\end{eqnarray*}
Then, using the equality
\begin{eqnarray*}
\int_0^\infty \frac{1}{1+u} u^{m+1}\mathrm{e}^{-\frac{\nu}{2\tau}u}du=\mathrm{e}^{\frac{\nu}{2\tau}}(m+1)!\Gamma\pg-1-m;\frac{\nu}{2\tau}\pd
\end{eqnarray*}
where $\Gamma(.;.)$ stands for the upper incomplete Gamma function, one obtains
\begin{eqnarray*}
I=C'\int_0^\infty \tau^{-\frac{\nu}{2}-m-2}\Gamma\pg-1-m;\frac{\nu}{2\tau}\pd d\tau
\end{eqnarray*}
where $C'=C(m+1)!\big(\frac{\nu}{2}\big)^{m+2}$. Since
\begin{eqnarray*}
\Gamma\pg-1-m;\frac{\nu}{2\tau}\pd=\pg\frac{\nu}{2\tau}\pd^{-m-1}E_{m+2}\pg\frac{\nu}{2\tau}\pd
\end{eqnarray*}
where $E_{m+2}$ is the generalized exponential integral, one has
\begin{eqnarray*}
I=C''\int_0^\infty \tau^{-\frac{\nu}{2}-1}E_{m+2}\pg\frac{\nu}{2\tau}\pd d\tau
\end{eqnarray*}
where $C''=C'\big(\frac{2}{\nu}\big)^{m+1}$ which leads to
\begin{eqnarray*}
I&=&C''\int_0^\infty \tau^{-\frac{\nu}{2}-1}\int_1^\infty \mathrm{e}^{-\frac{\nu}{2\tau}t}t^{-m-2} dt d\tau \notag \\
&=&C''\int_1^\infty t^{-m-2}\int_0^\infty  \tau^{-\frac{\nu}{2}-1} \mathrm{e}^{-\frac{\nu}{2\tau}t} dt d\tau \notag \\
&=&C'''\int_1^\infty t^{-m-2-\frac{\nu}{2}} dt=\frac{C'''}{m+1+\frac{\nu}{2}}
\end{eqnarray*}
where $C'''=C''\big(\frac{\nu}{2}\big)^{-\frac{\nu}{2}}\Gamma(\frac{\nu}{2})$ and finally
\begin{eqnarray*}
E[\psi(t_1)t_2]=\frac{(m+\frac{\nu}{2})m(m+1)}{m+1+\frac{\nu}{2}}.
\end{eqnarray*}
This leads to the following values for $a, b, c$ 
\begin{eqnarray*}
a&=&\frac{m(m+1)(m+\frac{\nu}{2})}{m+1+\frac{\nu}{2}}, \notag \\
b&=&\frac{m(m+1)(m+\frac{\nu}{2})}{m+1+\frac{\nu}{2}}, \notag \\
c&=&\frac{\nu}{2}\frac{m}{(m+1+\frac{\nu}{2})}+m^2 =\frac{m(m+1)(m+\frac{\nu}{2})}{m+1+\frac{\nu}{2}}.
\end{eqnarray*}
Substituting previous results in Eq.~\eqref{sigma-complex}, one finally obtains
\begin{equation*}
\sigma_1=\frac{1}{m+\nu/2} \quad \textrm{and} \quad \sigma_2=\frac{2(m+1+\nu/2)}{\nu(m+\nu/2)}.
\end{equation*}

\section{Proof of Theorem \ref{distance}}
\label{app-3}
To prove the statement of Theorem \ref{distance}, we will rewrite $\phi$ as 
\begin{equation}
\phi=(\phi_M-2\phi_{corr}+\phi_{SCM})/(\yb^H \Mb^{-1}\yb)^2
\end{equation}
where 
\begin{eqnarray*}
\phi_M&=&E\cg \pg f\pg\sigma\widehat \Mb\pd-f\pg\Mb\pd\pd ^2\cd \\
\phi_{corr}&=&E\cg \pg f\pg\sigma\widehat \Mb\pd-f\pg\Mb\pd\pd\pg f\pg\widehat \Mb_{SCM}\pd-f\pg\Mb\pd\pd\cd\\
\phi_{SCM}&=&E\cg\pg f\pg\widehat \Mb_{SCM}\pd-f\pg\Mb\pd\pd ^2\cd
\end{eqnarray*}
with $f(\Mb)=\yb^H \Mb^{-1}\yb$. Using the Delta method one can obtain  $\phi_M=f'(\Mb)\Sig_M f'(\Mb)^H$ \cite{huber1964robust} where $f'\pg\Mb\pd$ is the first derivative with respect to $\Mb$ and $\Sig_M=E\cg\pg\sigma\widehat \Mb-\Mb\pd\pg\sigma\widehat \Mb-\Mb\pd ^H\cd$ is defined by Eq. \eqref{asymp-M} (with $\Mb=\sigma\Mb_{\sigma}$). Moreover, one has a more general result $\phi_{corr}=f'\pg\Mb\pd\Sig_2 f'\pg\Mb\pd ^H=$ where $\Sig_2=E\cg\pg\sigma\widehat \Mb-\Mb\pd\pg\widehat \Mb_{SCM}-\Mb\pd ^H\cd=\gamma_1\pg\Mb^{T}\otimes \Mb\pd+\gamma_2\vec\pg \Mb\pd\vec\pg\Mb\pd^{H}$ with $\gamma_1$ and $\gamma_2$ the complex versions of Eq. \eqref{gamma}. In \cite{pascal2015asymptotic} it has been shown that $f'(\Mb)=\vec^H\pg\yb\yb^H\pd\pg\Mb^{T}\otimes \Mb\pd^{-1}$. Since $\pg\Mb^{T}\otimes \Mb\pd^{-1}\textrm{vec}\pg\Mb\pd=\textrm{vec}\pg\Mb^{-1}\pd$ one has
\begin{eqnarray*}
\phi_M&=&f'\pg\Mb\pd\Sig f'\pg\Mb\pd^H \\ \notag
&=&\vartheta_1 \textrm{vec}^H\pg\yb\yb^H\pd\pg\Mb^{T}\otimes \Mb\pd^{-1} \textrm{vec}\pg\yb\yb^H\pd\\ \notag
&+&\vartheta_2\textrm{vec}^H\pg\yb\yb^H\pd\textrm{vec}\pg \Mb^{-1}\pd\textrm{vec}\pg \Mb^{-1}\pd^{H}\textrm{vec}\pg\yb\yb^H\pd\\ \notag
&=&\vartheta_1 \textrm{vec}^H\pg\yb\yb^H\pd\textrm{vec}\pg\Mb^{-1}\yb\yb^H\Mb^{-1}\pd\\ \notag
&+&\vartheta_2\textrm{Tr}(\yb\yb^H\Mb^{-1})\textrm{Tr}( \Mb^{-1}\yb\yb^H)\\ \notag
&=&\vartheta_1 \textrm{Tr}(\yb^H \Mb^{-1}\yb\yb^H\Mb^{-1}\yb)+\vartheta_2 \textrm{Tr}(\yb^H\Mb^{-1}\yb)^2\\ \notag
&=&(\vartheta_1+\vartheta_2)(\yb^H \Mb^{-1}\yb)^2. \notag
\end{eqnarray*} 
It is now clear that $\phi_{SCM}=(\yb^H \Mb^{-1}\yb)^2$ and $\phi_{corr}=f'(\Mb)\Sig_2 f'(\Mb)^H=(\gamma_1+\gamma_2)(\yb^H \Mb^{-1}\yb)^2$ which leads to the final result.

\bibliographystyle{IEEEtran} 
\bibliography{biblio}

\end{document}